\documentclass[12pt]{article}

\usepackage{amsthm,graphicx,cite,lscape,float,multirow}
\usepackage{epsf,latexsym}
\usepackage{amssymb,amsmath}

\newcommand{\rhat}{\hat{\rm r}}

\newcommand{\nhat}{\hat{\rm n}}

\newcommand{\fp}{f_{\mbox{\scriptsize $\parallel$}}}

\begin{document}

\title{\bf On the lack of influence of contact area on the solid-liquid 
lateral retention force}

\author{Rafael de la 
Madrid,\footnote{E-mail: \texttt{rafael.delamadrid@lamar.edu}} \
Caleb Gregory,  
Huy Luong,\footnote{Current address: Johnson Controls, Port Arthur, TX 77640}
\ Tyler Stuck \\
\small{\it Department of Physics, Lamar University,
Beaumont, TX 77710} }

\date{\small{June 14, 2023}}









\maketitle

\begin{abstract}
\noindent We experimentally show that, unlike the solid-solid frictional 
force, the solid-liquid retention force is determined by interactions at 
the triple line rather than over the solid-liquid contact area,
as predicted by theory. We have prepared
drops whose triple line enclosed a uniform surface, a hydrophobic
island, and topographical inhomogeneities, and measured the same
retention force for all three cases. We 
have also studied the retention force on drops whose initial triple line was 
non-circumferential and measured different retention forces for different 
shapes, also as predicted by theory. The experiments with
non-circumferential drops provide (1) another way to show
that the retention force is not proportional to the contact area, and 
(2) a manner to falsify a recent
theory on drop depinning.

\end{abstract}

{\it Keywords}: Wetting; dewetting; adhesion; lateral retention force.

\section{Introduction}
\label{sec:intro}

Amontons' laws of friction state that the frictional force $f$ between 
two solid surfaces is proportional to the normal force (load) $F_{\rm N}$ and
is independent of the apparent contact area, $f=\mu F_{\rm N}$, where
$\mu$ is the friction coefficient~\cite{ISRA}. Amontons' laws, which hold 
under quite general conditions, seem to contradict the expectation that the 
frictional force should be proportional to the contact area. 


One of the main insights of the modern theory of friction~\cite{PT98,TABOR85} 
is that the frictional force is independent of the normal force 
and is in fact proportional to the actual contact area between the 
surfaces. This actual contact area arises from the points of actual 
contact between the surfaces, and it is much smaller than the apparent 
contact area. For rough surfaces, the actual contact area is proportional to 
the normal force, and therefore for rough surfaces the
frictional force is also proportional to the normal force, in agreement
with Amontons' laws~\cite{PT98,TABOR85}.

When a drop is trying to move on a solid surface, it experiences a force that 
opposes motion, called the lateral retention force, $\fp$, that is the 
solid-liquid analog of solid-solid friction. Because the liquid can deform, 
it seems natural to assume that the true contact area between the liquid 
and the solid is the same as the apparent contact area, and therefore one may 
be tempted
to conclude that the lateral retention force is proportional to the 
solid-liquid contact area, as suggested in Ref.~\cite{TADMOR21}.

The most widely accepted 
expression for the lateral retention force is given 
by~\cite{BROWN,DUSSAN,ANTONINI,CONINCK,DUNLOP}
\begin{equation}
    \fp = \gamma \oint_C \cos \theta\, \nhat \cdot \rhat \, ds  \, ,
        \label{netforcexin}
\end{equation}
where $\gamma$ is the liquid-vapor surface tension, $C$ denotes the triple line,
$\theta$ is the
(varying) contact angle along the triple line,  $\nhat$ is a unit vector 
perpendicular to the triple line, and $\rhat$ is a unit vector in the
direction of the overall motion of the drop. 

A feature of Eq.~(\ref{netforcexin}) that has not been highlighted in the 
literature is the fact that $\fp$ is independent of the properties
of the actual solid-liquid contact area, contrary to the naive 
expectation
that $\fp$ is proportional to such area~\cite{TADMOR21}. What is more, 
if we changed the chemical or topographical properties
of the contact area enclosed by the triple line, Eq.~(\ref{netforcexin})
predicts that $\fp$
remains unchanged as long as the contact angles remain unchanged. But, 
do contact angles remain unchanged when the properties of the contact area 
change?

For homogeneous surfaces, it has long been accepted that the contact angles
are not affected by the contact area (see, for example, 
Refs.~\cite{BROWN,RIO}). For heterogeneous surfaces, the pioneering studies 
of Wenzel~\cite{WENZEL36} and Cassie and Baxter~\cite{CASSIE44} seemingly 
established that contact angles are affected by the properties of the contact 
area enclosed by the triple line. However, a few 
studies~\cite{BARTELL53,EXTRAND03,MCCARTHY07,EXTRAND18} have shown that 
contact angles depend only on interactions at the triple line and are
unaffected by the properties contact area (see 
Ref.~\cite{ERBIL14} for a review on the debate on the dependence of 
contact angles on the triple line vs.~the solid-liquid contact area). Although 
in 2007 the minority view was that contact angles are unaffected
by the contact area~\cite{MCCARTHY07}, it seems that nowadays that has become
the majority view~\cite{ERBIL21}.

If the contact angles are unaffected by changes in the contact 
area enclosed by the triple line, Eq.~(\ref{netforcexin}) predicts that $\fp$ 
is also unaffected by those changes. Our first goal
is to provide experimental evidence that such is the case. Essentially, we will
set up experiments that are similar to those in 
Refs.~\cite{BARTELL53,EXTRAND03,MCCARTHY07,EXTRAND18}, but we will subject
the drop to an external centrifugal force and measure the force 
$\fp$ necessary to move the drop. We will show that at the onset of motion,
the lateral retention force is the same whether the
contact area is uniform, 
has a hydrophobic island, or has topographical defects.

It is well known that Eq.~(\ref{netforcexin}) predicts that $\fp$ depends
on the shape of the triple line. However, save for some exceptions
(see Refs.~\cite{AJP15,JANARDAN16,GRIEGOS2}), experimental
studies of the retention force usually deal with drops
whose initial triple line has a circumferential 
shape~\cite{EXTRAND,CARRE,ELSHERBINI,JANARDAN14,GRIEGOS1,TAFRESHI18d,RAFA19}.
Our second goal is to experimentally show that
different shapes of the triple line will in general lead to different
retention forces, as predicted by Eq.~(\ref{netforcexin}) and in agreement 
with the experimental results of Refs.~\cite{AJP15,JANARDAN16,GRIEGOS2}. We 
will prepare drops that are elongated 
along the parallel and perpendicular directions to the centrifugal force and 
show that their retention force is different from drops whose initial 
triple line is circumferential. The study of non-circumferential
triple lines will provide additional evidence that $\fp$ cannot be
proportional to the solid-liquid contact area.

When a drop is deposited at a given spot on a uniform, flat surface, its triple 
line has a circumferential shape. When an external
(e.g., centrifugal) force parallel to the surface is applied on it, the drop
starts leaning in the direction of
the force. Typically, as the external force increases,
the receding edge remains pinned to the surface, the advancing edge crawls 
forward, the drop spreads and, at the onset of motion, the shape of the
triple line somewhat resembles the egg of a hen. When the drop moves at low 
speeds (or, more precisely, when the capillary number is small), the shape of 
the triple line becomes ``quasi-rectangular''~\cite{FURMIDGE,RAFA22}. As the 
speed increases, the drop develops a tail that eventually
breaks up into small droplets~\cite{PODGORSKI}. Equation~(\ref{netforcexin}) does not describe this
equilibrium-to-motion transition, an additional theoretical input is needed to 
describe it. A recent theory~\cite{TADMOR22} describes some aspects of this 
transition. Specifically, the theory of Ref.~\cite{TADMOR22} asserts that,
in the equilibrium-to-motion transition, as the drop depins itself from the
surface to start moving,
the advancing edge will always move before the receding edge. Our third
goal is to show that our experiments with non-circumferential drops contradict
the theory of Ref.~\cite{TADMOR22} but are in agreement with the experiments 
of Ref.~\cite{JANARDAN16}.

Experimental verification of the consequences of Eq.~(\ref{netforcexin})
is important not just for its own sake, but also because Eq.~(\ref{netforcexin})
is not universally accepted as the correct expression for the lateral
retention force~\cite{TADMOR13,XU16,TADMOR21}. Our hope is that the
straightforward nature of our experiments will dispel any 
doubts~\cite{TADMOR13,XU16,TADMOR21} about the validity Eq.~(\ref{netforcexin}).

\section{Experimental Section}
\label{sec:expeapp}

\begin{figure}[ht!]
\begin{center}
              \epsfxsize=8.6cm
              \epsffile{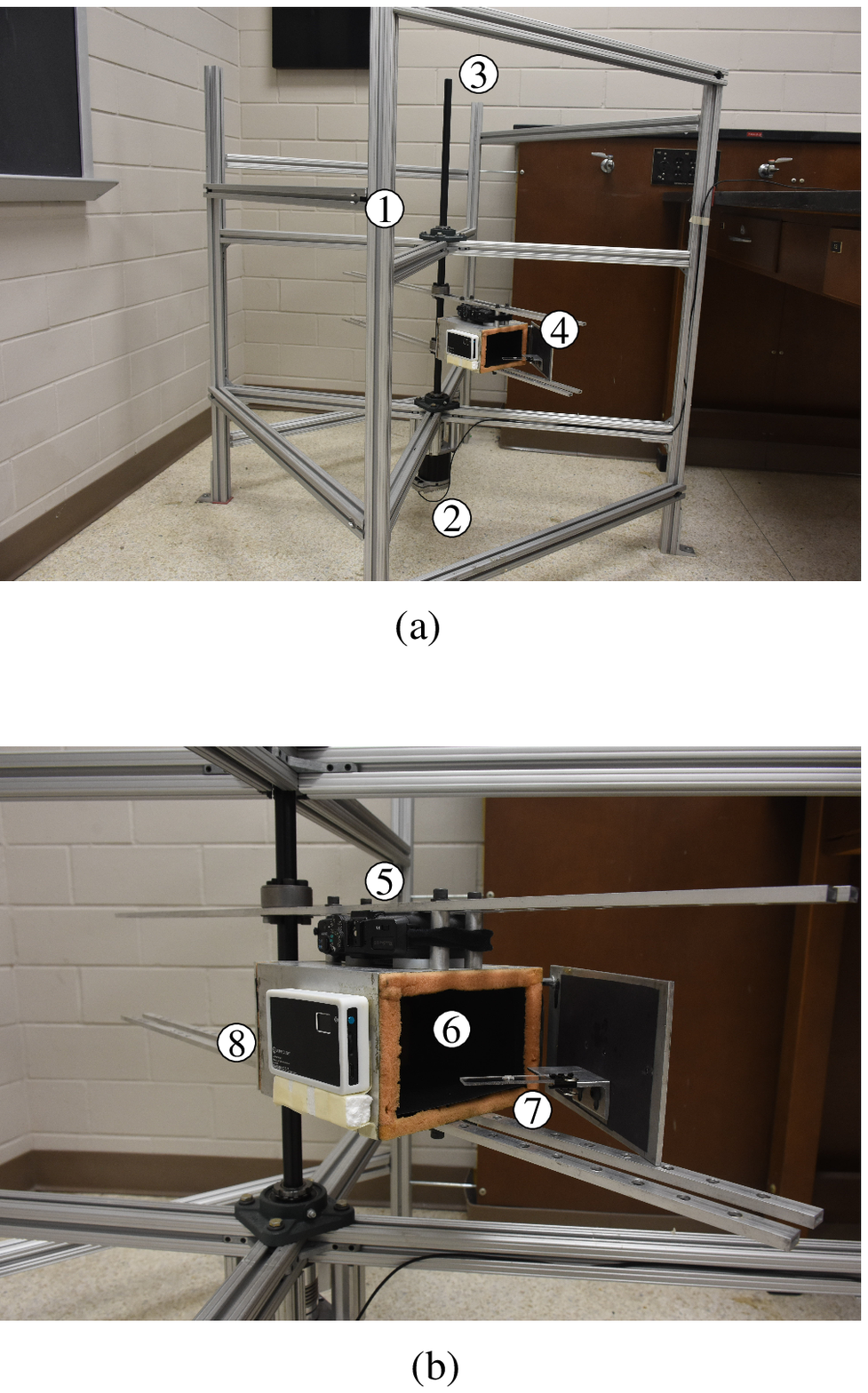}
\end{center}                
\caption{(a) Drop centrifuge: {\bf 1}-aluminum frame; {\bf 2}-motor; 
{\bf 3}-shaft; {\bf 4}-box. (b) Box: {\bf 5}-top camera; {\bf 6}-side camera;
{\bf 7}-PMMA sheet with drop; {\bf 8}-LED panel. Inside 
the box (not shown in the picture) there is a second LED panel that provides 
the necessary illumination for the top camera.}
\label{fig:apparatus}
\end{figure}

Our experimental apparatus 
is the same as in Ref.~\cite{RAFA19}, see Fig.~\ref{fig:apparatus}. It 
consists of a metallic frame with dimensions
$90~\text{cm} \times 90~\text{cm} \times 120~\text{cm}$ inside of which 
the rotary unit is mounted. The four legs of the frame are 
bolted to the floor to reduce mechanical vibrations. The rotary unit 
consists of a servo motor, a shaft, and two
pairs of aluminum rails that are attached perpendicularly to the shaft. The 
motor is connected to a power supply and a computer, whose software 
controls the motor. 

A metallic box is mounted on the aluminum rails. The box
has two cameras placed on top and on the side, which
provide top and side views of the drops. An 
LED panel mounted on one side of the box is used to set a common starting 
time in the videos of the cameras. On the door of the metallic box, we 
mounted a poly methyl methacrylate (PMMA) sheet (Optix$^{\tiny \textregistered}$, by 
Plaskolite) such that, when a drop is placed on 
the sheet and the door is closed, the cameras have side and top views of the 
drop. Illumination for
the side camera is provided by the outside LED panel. Lighting for the
top camera is provided by a second LED panel and an optical 
gradient~\cite{PODGORSKI}. The still frames of the
videos of the side (top) camera were fed into ImageJ~\cite{IMAGEJ} to measure
the contact angles (contact area and width) of the drops. We used
a syringe from Hamilton to deposit 60-$\mu$L drops of (distilled) water on 
the PMMA sheets. The large volume of the drops allowed us to build
drops whose initial triple line had a non-circumferential shape. 

\begin{figure}[ht!]
\begin{center}
              \epsfxsize=8.6cm
              \epsffile{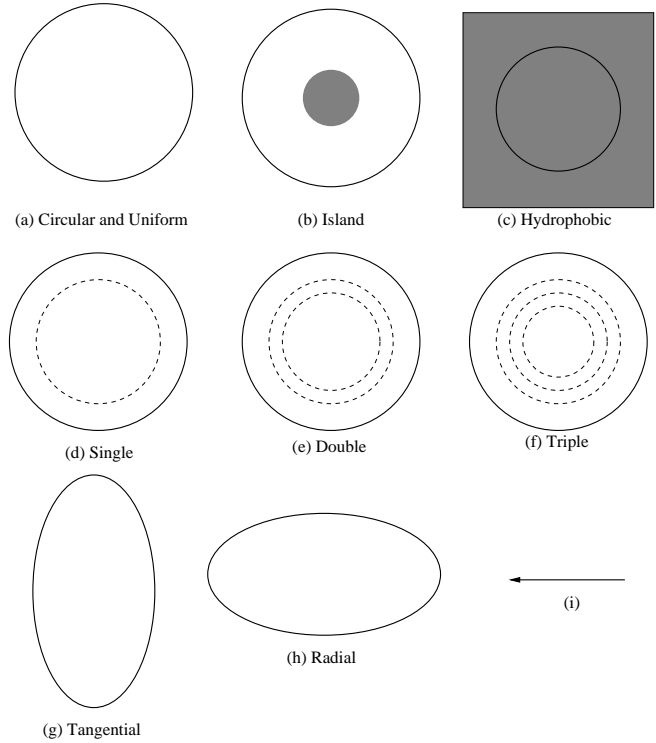}
\end{center}                
\caption{Schematic representation of the types of surfaces and shapes of
  triple line we used in this work. (a) Circular triple line (represented by
  a solid line) on a uniform surface; (b) 
  circular triple line encircling a hydrophobic island (represented
  by the gray circle); (c) circular triple
  line on a
  hydrophobic
  surface; (d) circular triple line encircling a single-circle scratch
  (represented by a dashed line);
  (e) circular triple line encircling a double-circle scratch; (f) circular
  triple line encircling a triple-circle scratch; (g) triple line
  elongated along
the direction perpendicular to the centrifugal force; (h) 
triple line elongated along
the direction of the centrifugal force; (i) direction of the
centrifugal force.}
\label{fig:shapes}
\end{figure}

Figure~\ref{fig:shapes} shows the different
surfaces and shapes of triple
lines used in this work. Figure~\ref{fig:shapes}(a) represents the
circumferential triple line of a drop resting on a uniform PMMA surface. Such
circumferential
triple line was obtained by slowly depositing the liquid on the same spot
of the surface. We
will refer to this kind of surface as ``uniform'' and this kind of triple
line as ``circular.'' Figure~\ref{fig:shapes}(b) represents a PMMA surface
that has a small hydrophobic island encircled by the triple line, hence
the name ``island.'' The 
hydrophobic island was 
obtained with Fusso Coat~\cite{SOFT99}. In the ``island'' case, as we
deposited the liquid on the center of the hydrophobic ``island,'' the
contact area of the drop increased, and eventually the
triple line moved into the PMMA surface and encircled the hydrophobic ``island''
completely. Figure~\ref{fig:shapes}(c) represents a hydrophobic surface 
obtained by covering the whole PMMA surface with Fusso Coat,
hence the name ``hydrophobic.'' When we
deposited the liquid at the same point on this hydrophobic surface, we obtained
a circumferential triple line, but with a smaller diameter than
in the ``uniform'' and ``island'' cases. Figure~\ref{fig:shapes}(d) represents
a PMMA surface with a 5-mm-diameter circular scratch etched with a laser
cutter. We will refer to this type of surface as ``single.'' When
a 60~$\mu$L drop
was deposited at the center of the circular
scratch, the triple line eventually moved past the scratch and formed a
circle of larger diameter than the scratch. The surfaces in
Figs.~\ref{fig:shapes}(e) and~\ref{fig:shapes}(f) are essentially the same
as in Fig.~\ref{fig:shapes}(d), but with the addition of a 3-mm-diameter 
(in Figure~\ref{fig:shapes}(e)) and a 1-mm-diameter (in 
Figure~\ref{fig:shapes}(f)) circular scratch. We will refer to those
surfaces as ``double'' and ``triple.''

On a uniform surface, besides the circular shape of 
Fig.~\ref{fig:shapes}(a), we built two other types of triple
line that have elongated shapes, along the
perpendicular (``tangential'') 
and parallel (``radial'') directions to the centrifugal force, see 
Figs.~\ref{fig:shapes}(g) and~(h). The direction of the centrifugal force
is represented by the arrow in Fig.~\ref{fig:shapes}(i). Those elongated drops
were built by depositing two small, 25~$\mu$L droplets close to each other. As
the two droplets grew, they eventually merged into a single drop. At the moment
of the merge, the single drop had an elongated, ``quasi-elliptical'' shape. To
keep the elongated shape, the remaining liquid in the syringe was
deposited at the ends of the ``major axis'' of the ``quasi-ellipse.''

The solid lines in Figs.~\ref{fig:shapes}(a)-(h) represent the initial shapes
of the triple line. When the centrifugal force starts acting on the drops,
those initial shapes start changing. For shapes (a)-(g), the advancing
edge starts moving first as the drop spreads (expands), whereas the
receding edge remains pinned at the surface. The onset of the motion is 
characterized by the instant when the receding edge starts moving. However,
for the ``radial'' shape of Fig.~\ref{fig:shapes}(h), the advancing edge
remains pinned to the surface, and the receding edge starts moving first as the 
drop contracts. The onset of the motion for ``radial''
drops is therefore determined by the instant when the advancing edge
starts moving.


The edge that determines the onset of the overall motion of the drop always
goes through a stick-slick
transition from rest to motion. There is therefore some ambiguity on 
when the actual onset of the motion occurs. Our criterion to 
determine the onset of the motion was that the
edge that determines it had to move at the most three pixels
(69~$\mu$m). We chose such criterion because after
the edge we were monitoring had moved three pixels, the stick-slick transition
was basically over. In some cases, the stick-slip transition was over after
the edge moved two or even just one pixel, and in such cases we used motion
of two or one pixel as the criterion to establish the onset of the 
motion. A three-pixel (69~$\mu$m) displacement 
is invisible
to the naked eye, and we used the custom software of Ref.~\cite{RAFA19}
to determine the instant at which the onset of the motion occurred. From
such instant, we calculated the centrifugal force at the onset of motion,
and from the centrifugal force we obtained the (maximum) lateral retention
force.

To accumulate enough statistics, for each case in Fig.~\ref{fig:shapes} we
prepared 15 different surfaces and did three runs on each surface. Hence,
the data for each case arises from about 45 runs. All the errors quoted
in this paper are statistical.

Although the exact integral expression of Eq.~(\ref{netforcexin}) has been used
in theoretical studies of the retention 
force~\cite{BROWN,DUSSAN,ANTONINI,CONINCK,DUNLOP},
in practical applications~\cite{AJP15,JANARDAN16,GRIEGOS2,EXTRAND,CARRE,ELSHERBINI,JANARDAN14,GRIEGOS1} 
the variation of the contact angle along the triple line 
is usually not known, and one approximates the three dimensional drop by a two
dimensional film. The triple line is then reduced to the advancing and
receding edges, and the exact integral expression of
$\fp$ is replaced by the following effective expression~\cite{DUSSAN,FURMIDGE}: 
\begin{equation}
    \fp =
     k \gamma w \left( \cos \theta _{\rm r}- \cos \theta _{\rm a} \right) ,
        \label{netforcex}
\end{equation}
where the shape factor $k$ is a fudge factor that
accounts for the unknown variation of $\theta$. In practice, one measures the 
retention force, the width, and the advancing and receding contact angles, 
whereas the shape factor is calculated as 
$k=\fp /w(\cos \theta _{\rm r}- \cos \theta _{\rm a})$. Therefore, we will 
be able to explain our experimental 
results in terms of Eq.~(\ref{netforcex}), but we will not be able to 
fully describe them using Eq.~(\ref{netforcexin}) due to our
inability to measure the contact angle along the triple line.

\section{Results and Discussion}

\subsection{First Experiment}

We first measured the lateral retention forces
on the ``uniform,'' ``island,'' and ``hydrophobic'' surfaces of 
Figs.~\ref{fig:shapes}(a)-(c).

\begin{table}[ht!]
\scalebox{0.9}{
\begin{tabular}{c| c |c| c| c| c| c| c| c c c c} 
  \hline
  \hline
Type  & $f_{\parallel}$ ($\mu$N) & $A_0$ (mm$^2$)
& $A$ (mm$^2$)
& $\theta_{\rm a}$$^{(\circ)}$
& $\theta_{\rm r}$$^{(\circ)}$ & $w$ (mm) & $k$ \\
\hline
\hline 
Uniform & $183\pm 4$ & $48.6\pm 0.2$ & $49.8 \pm 0.2$ &  $69.9 \pm 0.3$ & 
$47.2 \pm 0.7$ & $7.79 \pm 0.02$ & 
             $0.97 \pm 0.04$  \\ [1ex]
Island & $184 \pm 5$ & $48.1 \pm 0.3$ & $50.6 \pm 0.2 $ &  $69.2 \pm 0.4$ & 
$46.4 \pm 0.6$ & $7.78 \pm 0.02$ & 
          $0.98 \pm 0.04$  \\ [1ex]
Hydrophobic & $116 \pm 5$ & $29.4 \pm 0.3$  & $30.2 \pm 0.1$ & $101.5 \pm 0.6$ 
& $85.6 \pm 0.6$ & 
       $6.01 \pm 0.02 $  & $0.97 \pm 0.07$  \\ 
\hline
\hline 
\end{tabular}}
\caption{Experimental data of ``uniform,'' ``island,'' and 
``hydrophobic'' surfaces. }
\label{table:firstexperiment}
\end{table}

Table~\ref{table:firstexperiment} displays the magnitude of the lateral 
retention force $\fp$, the initial ($A_0$) and final $(A)$ 
contact areas, the advancing
$(\theta _{\rm a})$ and receding $(\theta _{\rm r})$ contact angles, the width $w$
of the drop at the onset of the motion, and the shape factor $k$. 

As can be seen in Table~\ref{table:firstexperiment}, the retention forces
of the ``uniform'' and ``island'' cases are roughly the same. Hence, the
onset of the motion of the drop is not affected by the
hydrophobic island. However, in the ``hydrophobic'' case, the triple line
sits on the hydrophobic coating and $\fp$ decreases dramatically. Thus, we
conclude that $\fp$ is not influenced by the hydrophobic island, and is
affected only when the
triple line interacts with the hydrophobic coating. We can also see in
Table~\ref{table:firstexperiment} that the other wetting parameters
(contact area, contact angles and width) are roughly the same in the
``uniform'' and ``island'' cases, but they change in the ``hydrophobic''
case, and therefore are affected by interactions at the triple line
rather than over the contact area. Hence,
Eqs.~(\ref{netforcexin}) and~(\ref{netforcex}) are consistent with our 
experimental
finding that $\fp$ is the same in the ``uniform'' and ``island'' 
cases, but changes in the ``hydrophobic'' one. Interestingly,
the shape factor of the ``hydrophobic'' case is very similar to those of the
``uniform'' and ``island'' cases, even though the other wetting parameters
are very different.

\subsection{Second Experiment}

In a second experiment, we compared the lateral 
retention forces on drops
deposited on the ``uniform,'' ``single,'' ``double,'' and ``triple'' surfaces
of Fig.~\ref{fig:shapes}, see data in Table~\ref{table:secondexperiment}.

\begin{table}[ht]
\scalebox{0.9}{
\begin{tabular}{c| c |c| c| c| c| c| c| c c c c} 
  \hline
  \hline
Type  & $f_{\parallel}$ ($\mu$N) & $A_0$ (mm$^2$)
& $A$ (mm$^2$)
& $\theta_{\rm a}$$^{(\circ)}$
& $\theta_{\rm r}$$^{(\circ)}$ & $w$ (mm) & $k$ \\
\hline
\hline 
Uniform & $187\pm 3$ & $48.6 \pm 0.1$ &  $50.8 \pm 0.1$ & $67.5 \pm 0.4$ & 
      $43.8 \pm 0.7$ &  $7.82 \pm 0.02$ & $0.98 \pm 0.04$  \\ [1ex]
Single & $182 \pm 3$ & $48.1 \pm 0.3$ &  $50.4 \pm  0.3$ & $69.4 \pm 0.5$ & 
         $46.4 \pm 0.7$ &  $7.75 \pm 0.04$ & $0.97 \pm 0.04$  \\ [1ex]
Double & $189 \pm 4$ & $48.5 \pm 0.2$ & $50.9 \pm 0.2$ & $68.9 \pm 0.3$ & 
       $44.1 \pm 0.4 $  & $7.76 \pm 0.02$  & $0.94 \pm 0.03$ \\ [1ex]
Triple & $180 \pm 5$ & $47.9 \pm 0.2$ & $49.7 \pm 0.2$ & $69.5 \pm 0.4$ & 
       $46.1 \pm 0.5 $  & $7.71 \pm 0.03$ & $0.94 \pm 0.04$  \\ 
\hline
\hline 
\end{tabular}}
\caption{Experimental data of ``uniform,'' ``single,'' 
  ``double,'' and ``triple'' surfaces.}
\label{table:secondexperiment}
\end{table}

The solid-solid frictional force certainly increases as the roughness of the
contact area increases. However, similar to our first experiment, the
retention force seems to be unaffected by the rough scratches contained
within the triple line, and has essentially the same value as we increase the
number of circular scratches from none (``uniform'') to three
(``triple''). As long as the triple line doesn't touch the
scratches, the motion of the drop remains unaffected by them. However, after
the motion of the drop starts, the triple
line eventually reaches the outer circular scratch, and the receding edge
gets pinned to the scratch. A centrifugal force of
about 330~$\mu$N is then necessary to depin the drop from the scratch. The
pinning
at the scratch is so strong that in many runs the drop breaks up and the
tail of the drop stays at the scratch while the rest of the drop moves
away. Thus, even though the drop is firmly pinned to the scratches when
the triple line reaches them, its motion is unaffected when the scratches are
contained within the triple line.

As can be seen in Table~\ref{table:secondexperiment}, the contact
angles, width and shape factor are essentially the same for all four types of
surface roughness, and therefore Eq.~(\ref{netforcexin}) 
and~(\ref{netforcex}) are consistent with our experimental findings.

\subsection{Third Experiment}

In the third experiment, we studied the influence of the shape of the
triple line on the retention force, by comparing the retention force of 
``circular,'' ``tangential,'' and ``radial'' drops on uniform PMMA
surfaces, see Fig.~\ref{fig:shapes}. Table~\ref{table:thirdexperiment}
displays the data of the third experiment.

  \begin{table}[ht]
\scalebox{0.9}{
    \begin{tabular}{c| c |c| c| c| c| c| c| c c c c} 
      \hline
      \hline
Type  & $f_{\parallel}$ ($\mu$N)\quad & $A_0$ (mm$^2$)
& $A$ (mm$^2$)
& $\theta_{\rm a}$$^{(\circ)}$
& $\theta_{\rm r}$$^{(\circ)}$ & $w$ (mm) & $k$ \\
\hline
\hline 
Circular & $187\pm 3$ & $48.6 \pm 0.1$ &  $50.8 \pm 0.1$ & $67.5 \pm 0.4$ & 
      $43.8 \pm 0.7$ &  $7.82 \pm 0.02$ & $0.98 \pm 0.04$  \\ [1ex]
Radial & $169 \pm 2$ & $52.9 \pm 0.3$ &  $50.6 \pm  0.1$ & $68.4\pm 0.3$ & 
          $44.5 \pm 0.5$ & $6.93 \pm 0.03$   &   $0.98 \pm 0.02 $  \\ [1ex]
Tangential & $215 \pm 3$ & $53.2 \pm 0.2$ & $53.9 \pm 0.4$ & $69.1 \pm 0.2$ & 
       $46.8 \pm 0.4$  & $9.46 \pm 0.05$   & $0.97 \pm 0.03$  \\ 
\hline
\hline 
    \end{tabular}}
\caption{Experimental data of drops with ``circular,'' ``radial'' and
``tangential'' triple-line shapes.} 
\label{table:thirdexperiment}
\end{table}

Contrary to solid-solid friction, the shape of the solid-liquid
contact area affects
$\fp$. The retention force for ``tangential'' drops is larger than for
``circular'' drops, which in turn is larger than for ``radial''
drops. Equation~(\ref{netforcex}) allows us to understand this result. Because
the advancing and receding contact angles and the shape
factor are very similar for the three shapes in this third experiment, what
basically determines $\fp$ is the width of the drop---the
longer the width, the larger the retention force. Because
$w_{\rm tangential}>w_{\rm circular}> w_{\rm radial}$, it makes sense that the
corresponding retention forces follow the same order.

As can be seen in 
Tables~\ref{table:firstexperiment}, \ref{table:secondexperiment}
and~\ref{table:thirdexperiment}, except for ``radial'' drops, the contact 
area at the onset of the motion is always larger than the initial one, that 
is, the drops spread before moving. Hence, as the retention force increases 
up to its maximum value, the contact area also increases up to its maximum 
value, and one may think that the size of the contact area may somehow 
affect the retention force, even though the roughness and the chemical 
composition of the contact area do not affect it. However,  
for ``radial'' drops the contact area at the onset of motion is
{\it smaller} than the initial one, see Table~\ref{table:thirdexperiment}. The 
reason is that for ``radial'' drops, as the
centrifugal drop increases, the advancing edge remains pinned to the surface
and the receding edge crawls forward, making the
contact area decrease. Hence, if the size of the contact area was an
essential factor affecting the lateral retention force, the contact area
of ``radial'' drops should also increase as the external centrifugal force
increases. However, for ``radial'' drops the opposite occurs.

As mentioned in the Introduction, Eq.~(\ref{netforcexin}) does not describe 
the changes that a drop experiences in the equilibrium-to-motion
transition, an additional theory is needed. Recently, a theory
to describe such transition was proposed in Ref.~\cite{TADMOR22}. According
to that theory, protrusion of the solid surface at the triple line always makes
the advancing edge of the drop move {\it before} the receding edge
does. However, our results for ``radial'' drops, in agreement with the
results of Ref.~\cite{JANARDAN16}, show that the
theory of Ref.~\cite{TADMOR22} is not correct, because for ``radial''
drops the advancing
edge moves {\it after} the receding edge.

\section{Conclusions}

We have presented three experiments that show 
that the lateral retention
force is determined by the interactions at the triple line and by the 
shape of the
triple line itself, and seems to be unaffected by the properties of the contact
area enclosed by the triple line, in contrast
to the solid-solid frictional force, which depends on the size and 
roughness of the
actual contact area but does not depend of the shape of its perimeter. If
interactions
over the contact area do affect $\fp$, their influence is much smaller than 
the influence of the
interactions at the triple line. From a theoretical point of view, our
experiments are a straightforward consequence of Eq.~(\ref{netforcexin}), which
provides an expression for the lateral retention force that depends only on
quantities defined at the triple line and is unaffected by
the contact area. 

As expected~\cite{MCCARTHY07}, we have found that the advancing and receding
contact angles are independent of the initial shape of the triple line, and
of the chemical and topographical properties of the contact area. Somewhat
unexpectedly, we have found that the shape factor is very similar in all
cases, even for different shapes of the triple line.

Our results are consistent with, and expand the results of
Refs.~\cite{BARTELL53,EXTRAND03,MCCARTHY07,EXTRAND18,JANARDAN16,GRIEGOS2},
but they are in disagreement with some of the conclusions of
Refs.~\cite{TADMOR13,XU16,TADMOR21,TADMOR22}. In particular, our experiments
with non-circumferential triple lines are in disagreement with the
theory of Ref.~\cite{TADMOR22} on drop depinning.

Finally, we would
like to note that, although wetting quantities such as contact
angles and retention forces depend only on interactions at the triple line,
some energy quantities do depend on contact area. Specifically, the energies per
unit of contact area to expand, contract, slide or detach a drop are
constant, and do not depend on the shape of the triple
line~\cite{RAFA22}.

\section*{CRediT authorship contribution statement}

Rafael de la Madrid: Conceptualization, data acquisition, writing-review \&
editing. Caleb Gregory: Data analysis. Huy Luong: Design and construction of 
experimental apparatus. Tyler Stuck: Data analysis.

\section*{Acknowledgments}

Financial support from Lamar COAS and SURF 
fellowships is gratefully acknowledged.

\newpage

\newpage

\newpage

\begin{figure}[ht!]
\begin{center}
              \epsfxsize=15cm
              \epsffile{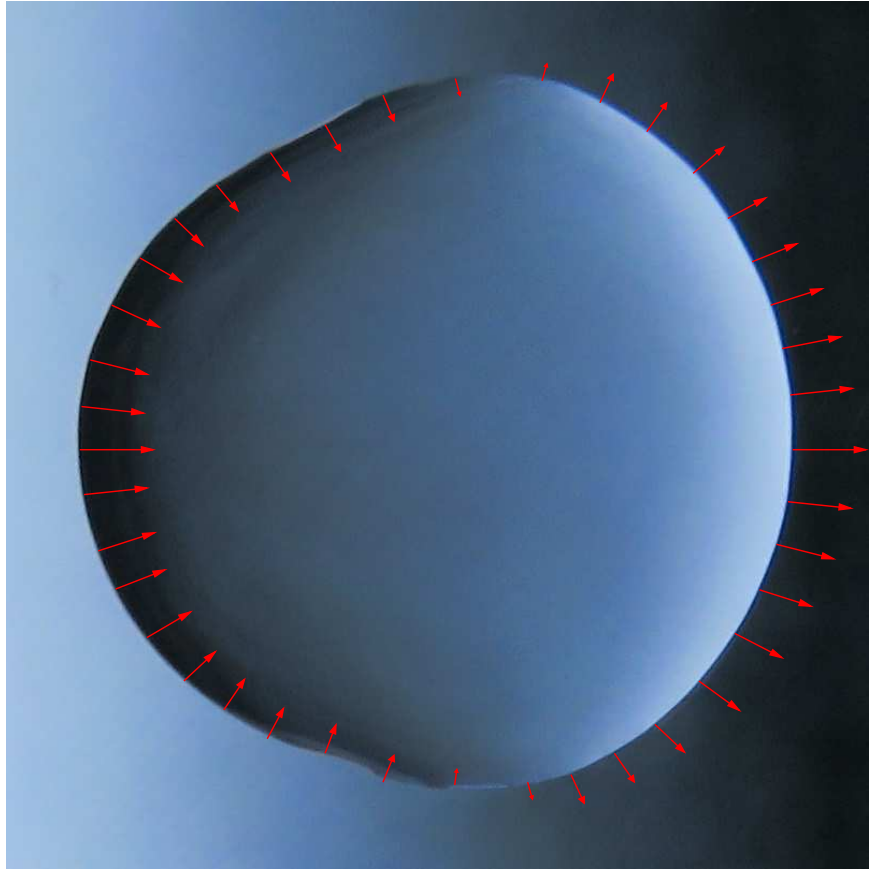}
\end{center}                
\caption{Graphical abstract.}
\label{fig:toc}
\end{figure}

\end{document}